\documentclass[prb,twocolumn,floatfix, showpacs,showkeys]{revtex4}

\usepackage{verbatim}
\usepackage{epsfig}

\newcommand{\nod}{\noindent}

\usepackage{subfigure}

\usepackage{amsmath}

\usepackage{natbib}

\begin{document}

\title{Simulation of Cu-Mg metallic glass: Thermodynamics and Structure}

\date{\today}












\author{Nicholas P. Bailey}
\email{nbailey@fysik.dtu.dk}
\affiliation{CAMP, Department of Physics, Technical University of Denmark, 
DK-2800, Lyngby, Denmark and Materials Research Department, Ris{\o} National 
Laboratory, DK-4000, Roskilde, Denmark}

\author{Jakob Schi{\o}tz}
\author{ Karsten W. Jacobsen}
\affiliation{CAMP, Department of Physics, Technical University of Denmark, 
DK-2800 Lyngby, Denmark}

\begin{abstract}
We have obtained  effective medium theory (EMT) interatomic potential 
parameters suitable for studying Cu-Mg metallic glasses. We present 
thermodynamic and structural results from simulations of such glasses over a 
range of compositions. We have produced low-temperature configurations by 
cooling from the melt at as slow a rate as practical, using constant 
temperature and pressure molecular dynamics. During the cooling process we have
carried out thermodynamic analyses based on the temperature dependence of the
enthalpy and its derivative, the specific heat, from which the glass transition
temperature may be determined. We have also carried out structural analyses
 using the radial distribution function (RDF) 
and common neighbor analysis (CNA). Our analysis suggests that the splitting of
 the second peak, commonly
associated with metallic glasses, in fact has little to do with the glass
transition itself, but is simply a consequence of the narrowing of peaks 
associated with structural features present in the liquid state. In fact the 
splitting temperature for the Cu-Cu RDF is well above $T_g$. The CNA also
highlights a strong similarity between the structure of the intermetallic 
alloys and the amorphous alloys of similar composition. We have also 
investigated the diffusivity in the supercooled regime. Its temperature 
dependence indicates fragile-liquid behavior, typical of binary metallic 
glasses. On the other hand, the relatively low specific heat jump of around 
$1.5 k_B/\mathrm{at.}$ indicates apparent  strong-liquid behavior, but this can
 be explained by the width of the transition due to the high cooling rates.
\end{abstract}

\maketitle

\section{\label{introduction}Introduction}


Metallic glasses~\cite{Greer:1995, Johnson:1999} have generated considerable
 scientific interest since they were discovered 40 years ago, due to their 
unusual magnetic and mechanical properties, as well as wear and corrosion 
resistance, and their glass-forming ability {\it per se}. This interest has 
substantially increased since the discovery of the so-called bulk metallic
 glasses (BMGs) or bulk amorphous alloys, by Inoue~\cite{Inoue/others:1991}
 and Johnson.\cite{Peker/Johnson:1993} The ability to create samples with 
thicknesses in the mm or cm range, rather than $\mu$m thick ribbons, greatly
 increases the applicability of the materials, as well as the range of 
measurements that can be performed on them. This is particularly true in the
 case of mechanical testing, and recently measurements of properties such as
 fracture toughness, fracture morphology and crack-tip plasticity have been
 made.\cite{Fedorov/Ushakov:2001,Flores/Dauskardt:2001, 
Schneibel/Horton/Munroe:2001, Gilbert/Ritchie/Johnson:1997}


The mechanisms of plastic deformation are of particular interest in metallic 
glasses in view of the fact that there are no obvious topological defects 
which might play a role analogous to crystal-dislocations, allowing slip to
 take place in small increments. Thus metallic glasses tend to have very high 
flow stresses.\cite{Greer:1995}  A complete understanding of plastic 
deformation must include the following two parts: (i) detailed knowledge of the
 elementary events that constitute plastic flow and (ii) a
practical continuum theory which uses this knowledge to make predictions of
macroscopic behavior (a recent such theory is the so-called Shear 
Transformation Zone (STZ) 
theory\cite{Falk/Langer:1998, Langer/Pechenik:2003}). The motivation for the 
present work is a desire to tackle item (i) using the tools of modern materials
 simulations, specifically: realistic potentials, system sizes as large as 
feasible and necessary, and sophisticated analysis and visualization 
techniques. The first step, addressed in this paper, is to create appropriate
interatomic potentials, generate glassy configurations, and study the
 thermodynamics and structure of the system, in order to understand it as a 
glass-forming one. Simulations of mechanical properties will be presented in
future publications. The phrase ``realistic potentials'' refers to 
contemporary potentials commonly used for metals, 
including effective medium or embedded atom-type potentials, or
 pseudopotential-based pair potentials, as opposed to Lennard-Jones potentials,
 which are commonly used (with two components) to model metallic 
glasses.~\cite{Kob/Andersen:1995a, Kob/Andersen:1995b,Sastry/Debenedetti/Stillinger:1998,Utz/Debenedetti/Stillinger:2000,Lacks:2001,Varnik/others:2003} Such 
potentials are especially useful because they allow quantitative comparison
with experiments of properties such as glass transition temperature, and, 
later, mechanical properties.


In this paper we present molecular dynamics simulations of the binary alloy 
Cu$_x$Mg$_{1-x}$. Mg-based BMGs such as 
Mg$_{60}$Cu$_{30}$Y$_{10}$\cite{Inoue/others:1991, Busch/Liu/Johnson:1998, 
Wert/Pryds/Zhang:2001, Wolff/Pryds/Wert:2003}  are of interest because 
their weight is low, being dominated by Mg, but their strength can be 
comparable to high-strength steel. We have chosen to study the binary alloy 
because (i) it is simpler to optimize a potential for two species than for 
three and (ii) it is easier to study dependence on a single 
composition-parameter than on two. Our intent to use realistic potentials 
necessitates an attempt to create as
realistic a glass as possible with those potentials. It is thus important to 
characterize the system as a glass-forming and alloying one as completely as
 possible. 

\begin{figure}
\epsfig{file=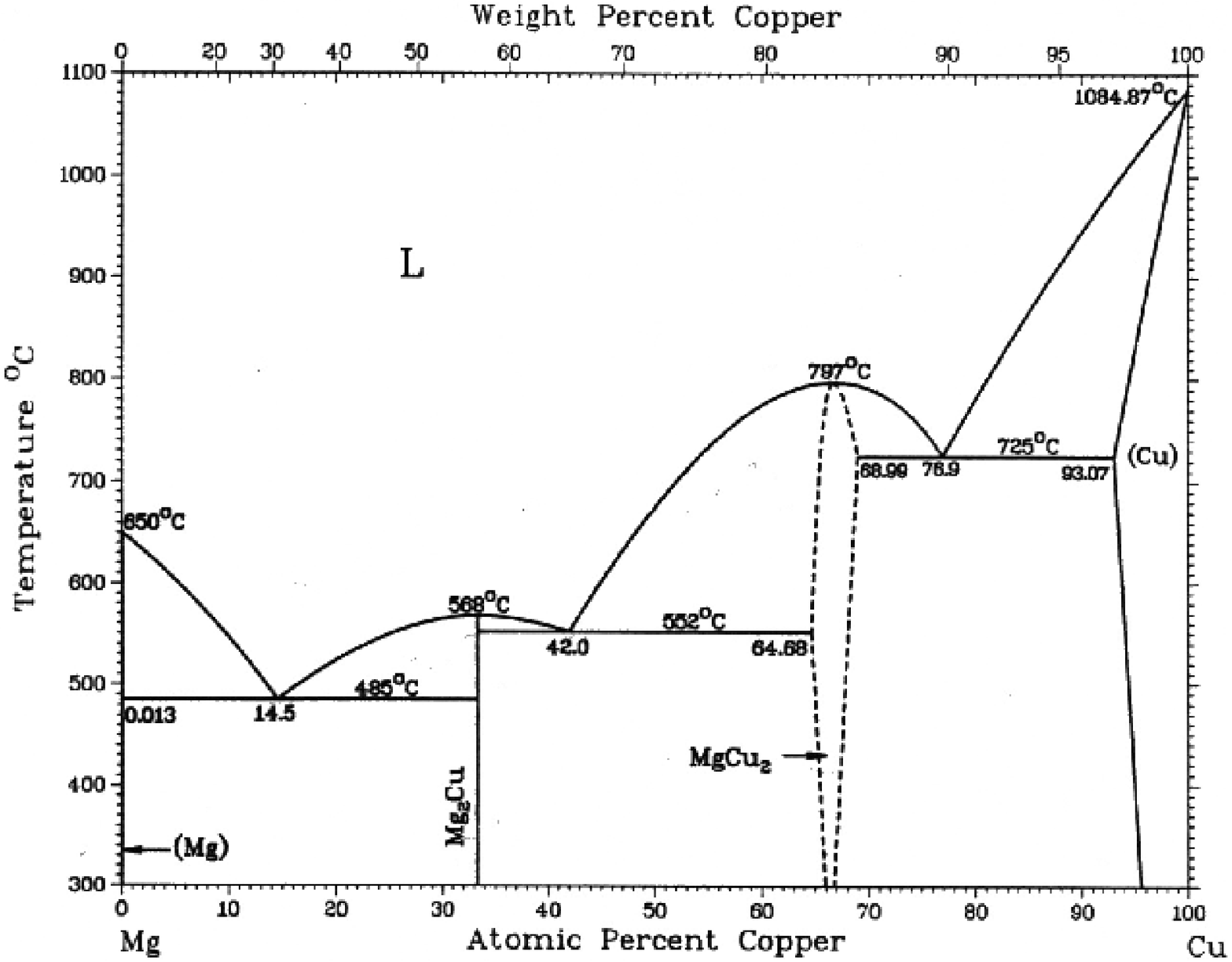,width=3.3 in}
\caption{\label{CuMgPhaseDiagram}Equilibrium phase diagram for Mg-Cu (from 
Ref.~\onlinecite{BinaryAlloyPhaseDiagrams})}
\end{figure}

 The Cu-Mg equilibrium phase diagram is shown in Fig.~\ref{CuMgPhaseDiagram}. 
Experimentally it forms a glass over a range of compositions 
from 9--42 at. \% (complete glass formation over 12--22 \%, which includes the
 eutectic composition 14.5 \%).\cite{Sommer/Bucher/Predal:1980} It is not a 
BMG, since it can only be formed by melt-spinning at high cooling rates. The  
cooling rates in the simulations are necessarily even higher and allow glassy 
configurations to be created over almost the entire range of compositions. 
It is worth studying the experimentally inaccessible states as part of the 
process of detecting trends in material properties as a function of 
composition; it is the crystal-nucleation timescale, lying between the 
simulation and experimental timescales, which makes the difference between
crystalline and amorphous phases---if just a few orders of magnitude gain in
cooling rate could be experimentally realized, there is reason to believe these
states would be as stable as the actual 
glassy configurations currently realizable by experiment.

Because the crystallization rates are high, there are limited experimental 
measurements of the thermodynamic properties of Mg-Cu glasses, and thus it
is of interest to study these in the simulations before moving on to 
mechanical properties. In the process we find some interesting results 
regarding  structural changes in the super cooled regime (steady growth of 
icosahedral order and evidence of restructuring thermodynamics). Additionally 
we  make some observations on the question of the fragility of this system.
The next section will discuss some aspects of the theory of glass formation
in alloys, as applied to the Cu-Mg system. Section~\ref{simulationMethods} will
 discuss simulation methods, including the fitting of the interatomic 
potential. 
Sections~\ref{glassTransition} and~\ref{structuralAnalysis} 
discuss characterization of the glass transition and of structural properties
respectively. The last section is the discussion.

\section{Glass formation in the M\lowercase{g}-C\lowercase{u} system}





One approach to the theory of metallic glass formation is based on
 pseudopotential-derived interatomic 
pair potentials,\cite{Hafner:1980, Hafner:1987} and emphasizes the coincidence
 of bond 
lengths with potential minima. We will not be using such potentials; in fact
many aspects of glass formation are 
purely geometrical (packing of spheres) and phase-energetic\cite{Hafner:1980} 
(comparison with competing crystalline phases).
Frank and Kasper~\cite{Frank/Kasper:1958,Frank/Kasper:1959} pointed out that
many complex intermetallic structures can be understood in terms of tetrahedral
close-packing of spheres. Examples of so-called Frank-Kasper (FK) 
phases include the Laves phases (C14, C15, C36) and $\mu$, $\chi$
and $\sigma$ phases. The high packing fraction and coordination numbers 
suggest directional bonding does not play a role. The closest packing of 
spheres of equal size is achieved with a tetrahedron (79\%), but tetrahedra 
cannot fill space---the best one can do is make an icosahedron out of twenty 
slightly distorted tetrahedra, but this cannot be repeated periodically, so 
in crystals one has the fcc (e.g. Cu, $a=3.61$\AA) and hcp (e.g. Mg, 
$a=3.21$\AA, $c=5.21$\AA) structures, with 74\% packing.

In the Cu-Mg system there is indeed a Laves 
phase, Cu$_2$Mg. This is not
surprising given that the ideal Laves packing is achieved with a radius ratio
of 1.225 (Ref.~\onlinecite{Hafner:1987}, p.59), which is close to that of Mg
and Cu (1.256 using the Goldschmidt radii, based on nearest neighbor distances
 of the pure metals). This 
phase is quite stable simply because having a majority of smaller atoms allows
 a greater packing fraction. On the other hand Mg$_2$Cu, with the larger atoms
 in the majority, is not as stable an alloy.\cite{Hafner:1980} Mg-Cu is in a 
class of metallic glass formers which include simple metal-transition metal
binary alloys and are characterized by a Laves phase when the small atom (Cu) 
is in the majority, and a glass when the larger atom is. In Cu$_2$Mg the Cu 
atoms have CN12 icosahedral coordination and the Mg atoms are 16-coordinated, 
surrounded by so-called Frank-Kasper 16-hedra (more specifically, Friauf 
polyhedra).

Glass formation in a binary alloy appears to be favored by the same criteria
that favor the formation of FK phases: large, negative heats of 
formation, non-directionality of bonding, and a tendency to maximize packing 
fraction. In general one finds that for compositions between intermetallics
(for example near eutectics), where the equilibrium phase diagram shows a
 two-phase mixture, the amorphous phase is more stable than any single 
crystalline phase. In the region of the phase diagram where FK phases appear, 
glass formation typically loses out in the competition experimentally,
presumably because the nucleation of the Laves phase is rather easy. In the 
Cu-Mg system, the region of experimental glass formation is on the Mg-rich 
side, where the competing crystalline phase, Mg$_2$Cu, is quite complex 
(48 atoms in the unit cell).

\section{\label{simulationMethods}Simulation methods}

\subsection{\label{potentials}Potentials}



The interatomic potential we use is the effective medium theory 
(EMT),\cite{Jacobsen/Norskov/Puska:1987,Jacobsen/Stoltze/Norskov:1996} fit to data obtained from density 
functional theory (DFT) calculations and 
experiment. This has previously been applied to fcc metals,
in particular late transition and noble metals and has been of great use in
 studying mechanical properties of crystalline 
metals.\cite{Schiotz/others:1999, Schiotz/Jacobsen:2003} As Mg crystallizes in 
hcp with an almost ideal $c/a$-ratio of $1.624$ (ideal is 
$\sqrt{8/3}=1.633$), indicating little directional bonding, we might expect it 
to be reasonably well described by an appropriately optimized EMT potential.

EMT uses seven parameters for each element. A set of parameters for Cu exists 
but these have been optimized for simulations of pure, crystalline Cu (where 
for example particular attention was paid to the stacking fault energy, which
 is of no concern in amorphous materials). For the amorphous alloys, it is 
important that the formation
energies are reasonable, in particular that they are negative (otherwise the
system will simply separate into regions of pure Cu and regions of pure Mg).

Thus we have (re-)fit the parameters of both elements, taking into account
basic properties such as lattice constants, cohesive energies and elastic 
constants of the pure 
elements, as well as the formation energies of the two 
intermetallic compounds, Mg$_2$Cu and Cu$_2$Mg. Due to the near-ideal hcp 
packing of Mg, its 
structure differs from fcc only at the second neighbor level. For simplicity, 
and because the EMT potential is formulated in terms of fcc packing, we used
calculated properties of fcc Mg in the fitting, except that the cohesive 
energy was
corrected using the experimental hcp value and the calculated fcc-hcp 
difference ($23 meV$/atom), calculated differences in cohesive energy being 
expected to be more accurate than calculated cohesive energies themselves.

\begin{table}

\begin{tabular}{|c|c|c|}
\hline
  parameter & Cu & Mg \\
\hline
$s_0$ & 2.67 &  1.766399 \\	
$E_0$ & -3.51 &  -1.487 \\
$\lambda$ & 3.693666 & 3.292725	\\
$\kappa$ & 4.943848 & 4.435425 \\
$V_0$ & 1.993953 & 2.229870 \\	
$n_0$ & 0.063738 & 0.035544 \\	
$\eta_2$ &  3.039871 & 2.541137 \\
\hline	
\end{tabular}
\caption{\label{EMTparameters}EMT parameters for Cu and Mg, in units derived
 from $eV$ and \AA. }
\end{table}

\begin{table}
\begin{tabular}{|c|c|c|}
\hline
property & optimized value & target value \\
\hline
Cu-$E_{\textrm{coh}}$ & 3.521 & 3.510 \\
Cu-$a$ & 3.588 & 3.610 \\
Cu-$B$ & 0.891 & 0.886 \\
Cu-$C_{44}$ & 0.512 & 0.511 \\
Cu-$C_{11}$ & 1.095 & 1.100 \\
Mg-$E_{\textrm{coh}}$ & 1.487 & 1.487 \\
Mg-$a$ & 4.502 & 4.520 \\
Mg-$B$ & 0.242 & 0.225 \\
Mg-$C_{44}$ & 0.117 & 0.115 \\
Mg-$C_{11}$ & 0.293 & 0.326 \\
Mg$_2$Cu-$\Delta H$ & -0.115 & -0.132 \\
Mg$_2$Cu-$a$ & 5.250 & 5.320 \\
Cu$_2$Mg-$\Delta H$ & -0.159 & -0.157 \\
Cu$_2$Mg-$a$ & 6.943 & 7.158 \\
\hline
\end{tabular}
\caption{\label{fitProps}Properties used in the fitting: the values specified
 (from DFT/experiment) and the values according to the optimized potential. 
$B=$ bulk modulus, $a=$ lattice constant.}
\end{table}

The optimized EMT 
coefficients are shown in table~\ref{EMTparameters} and the target and fitted 
values of the fitting properties are shown in table~\ref{fitProps}. Note that
for the orthorhombic Mg$_2$Cu, the experimental $b/a$ and $c/a$ were used, as 
well as the experimental values of the internal coordinates. The alloy 
formation energies are well-represented. Unlike pair potentials based 
upon pseudopotentials, the present form of the 
EMT potential\cite{Jacobsen/Stoltze/Norskov:1996} does not incorporate the 
Friedel oscillations, and the idea that stability of intermetallic compounds is
 determined by the matching of minima of pair potentials to interatomic 
distances\cite{Hafner:1980} does not play a role; the fact that EMT
parameters can be chosen to give the correct formation energies of the 
intermetallic compounds appears to be most important.

\subsection{\label{molecularDynamics}Molecular dynamics}

We simulated the cooling of systems of 2048 atoms from the liquid state (above
the melting point) down to
 zero temperature. The compositions ranged from
pure Mg to pure Cu, and are labeled by the percentage of Cu. For most
simulations we used 21 compositions, increasing in steps of 5\% from 0 (pure
Mg). The initial state was an fcc lattice with the sites occupied at random by 
Cu or Mg atoms in accordance with the overall composition. There was no initial
heating phase; the first stage in the cooling run set the temperature to a 
value well above the melting point (values ranged from 1392 K for Mg 
($T_m = 923 K$) to 1857 K for Cu ($T_m = 1358 K$)), making the crystal
melt immediately. Two rates of cooling were
 used; differing in the amount of simulation 
time at each temperature. Cooling took place in steps of $35K$; 
the procedure at each temperature stage was as follows: (i) a small number of 
steps, corresponding to $0.6ps$ (the MD time-step was $2fs$), of
constant-volume Langevin thermalization was carried out in order to 
approximately thermalize the system to the new temperature; (ii) the dynamics
was switched to constant-pressure (NPT) dynamics and the system was simulated
for an initial equilibration time of $6 ps/ 12 ps$; (iii) the system was 
simulated for a longer time $40 ps/120 ps$ during which thermal averages of 
various quantities of interest were taken. This time also contributed to the 
equilibration of the system. The overall cooling rates were thus close to 
$0.72 K/ps$ ($7.2 \times 10^{11}K/s$) and $0.25 K/ps$ 
($2.5 \times 10^{11}K/s$). The NPT dynamics used was a combination of 
Nose-Hoover and Parrinello-Rahman dynamics, proposed by
 Melchionna.\cite{Melchionna/Ciccotti/Holian:1993, Melchionna:2000, 
Holian/others:1990, Tolla/Ronchetti:1993} We turned off shearing, allowing only
 volume fluctuations, because the
liquid state cannot support a shear stress and fluctuations in the periodic
box sometimes lead to extreme angles between box vectors and thus problems with
the neighbor-locating algorithm. The pressure was zero or a 
small positive value (this was necessary in some cases when the initial 
temperature was above the boiling point of pure Mg). For each cooling rate the
simulations were run twice with different random number seeds (affecting the
distribution of species in the initial lattice and the Langevin dynamics used
when the temperature is changed; the NPT dynamics does not use random numbers).


During the averaging period, the pressure, volume, kinetic and 
potential energies were recorded and averaged. For the purposes of structural
analyses so too was the radial distribution function (RDF), both total and 
separated into contributions from Mg-Mg, Cu-Cu and Mg-Cu. At the end of the 
averaging time
the current configuration was saved, as well as a configuration obtained from 
it by direct minimization (quenching) using the MDmin minimization algorithm.
At a later time the saved configurations from selected temperatures were used
for further simulation at that temperature to gather further dynamical and
structural information such as diffusion constants and thermally averaged 
common neighbor analysis (CNA).

\begin{figure}
\epsfig{file=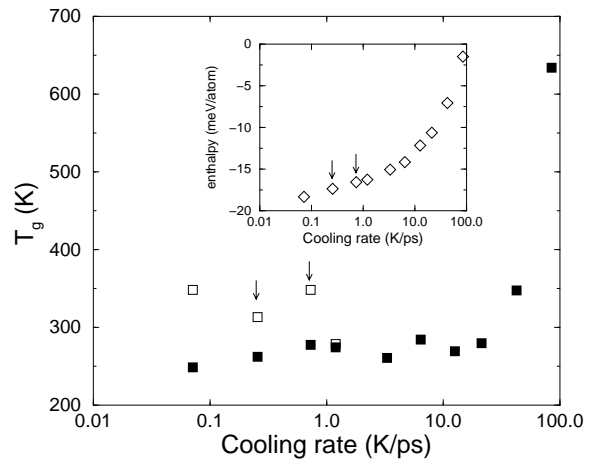, width = 3.0 in}
\caption{\label{CoolingRateFig}Cooling rate dependence of $T_g$ for 15\% Cu
system. Solid symbols, $T_g$ from the maximum rate of change of $C_P$;
 open symbols, $T_g$ by intercept method. Arrows indicate the cooling rates
 used for the main simulations. Inset: enthalpy of system at
end of cooling run ($T=0$).}
\end{figure}

Our cooling rates are as slow as in other recent simulations of amorphous 
metals,\cite{Hui/others:2002, Qi/others:1999,Posada-Amarillas/Garzon:1996,
Lewis:1989} but they are of course larger than experimental rates by several 
orders of magnitude. In order to check that our results are not significantly 
affected by this difference, we have cooled one composition, 15\% Cu at several
 faster rates and one slower one. Fig.~\ref{CoolingRateFig} shows $T_g$ and the
 enthalpy at $T=0$ for these runs. The methods of calculating $T_g$ are 
explained
 in the next section; only one (intercept) could be used for the very 
fast runs. It is pretty clear that for the cooling rates used in the main 
simulation, the dependence of $T_g$ on cooling rate has become smaller than the
 uncertainty in determining $T_g$. The enthalpy
shows a definite slope still at the lowest cooling rate, amounting to about 
$1meV$ per order of magnitude cooling rate, which is rather small; also one 
would expect the curve to flatten out more at even smaller rates. The one
significant difference we notice is that crystallization at the Cu-rich end 
happens at lower Cu-concentrations for slower cooling: the 90\% Cu system
crystallizes in one run at $0.25 K/ps$ but not at all at $0.72 K/ps$.

\section{\label{glassTransition}Glass transition}


\begin{figure}
\epsfig{file=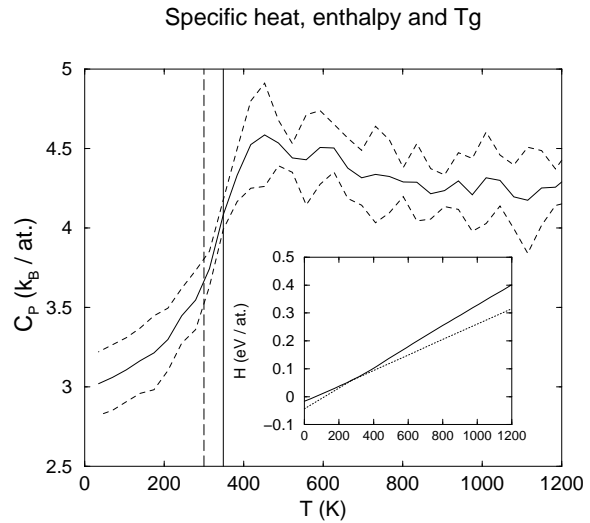,width=3.0 in}
\caption{\label{specHeatEnth}Specific heat versus temperature for 15\% Cu 
system, cooling rate $0.72 K/ps$. Dashed lines, values from two separate 
cooling runs, displaced $\pm 0.2$ for clarity. Solid line, average of these 
two. Solid vertical line, $T_g$ from maximum slope of specific heat; solid 
dashed line, $T_g$ from intercept method. Inset: enthalpy (average
 of the two runs) versus temperature. Dotted lines show the extrapolated 
straight-line fits from the intercept method.}
\end{figure}

We see glass transitions in almost all compositions, the exceptions being the
pure elements and 95\% Cu, which crystallize in fcc/hcp 
structures (also 90\% Cu in one out of two runs at $0.25 K/ps$).
The first evidence that a glass transition takes place upon cooling appears in
the enthalpy versus temperature curve, which shows a change in slope (inset in
 Fig~\ref{specHeatEnth}). This suggests a way to determine $T_g$ by breaking 
the curve into two pieces, fitting a straight line to each, and intersecting 
the two lines obtained. We call this the ``intercept-method''. It turns out 
that this tends to underestimate $T_g$ as can be seen by looking at the 
derivative of the enthalpy, the specific heat (Fig~\ref{specHeatEnth}), 
obtained from centered differences. The
$T_g$ ends up at the leftmost part of the steep part of the curve, whereas one
would expect any reasonable definition of $T_g$ to be roughly in the center of
the transition region (defined as the steep part). Thus we compute $T_g$ as the
temperature at which the specific heat is changing fastest by taking
derivatives again and simply choosing the maximum. This method necessarily
yields a $T_g$ equal to one of the simulation temperatures, but since the
transition region is a few times wider ($150-200 K$) than the temperature step 
in the simulation, one cannot expect to do better (experimentally one sees
widths of some tens of $K$, see for example 
Ref.~\onlinecite{Busch/Liu/Johnson:1995}). In cases where we have two different
enthalpy curves for the same system cooled identically but from different 
starting configurations we average the two enthalpy curves before applying the
method, as this gives a smoother $c_P$ curve.

The $T_g$ we get for 15\% Cu is 
$350 K$ which is remarkably similar to the experimental value of $380 K$ 
reported by Sommer et al.\cite{Sommer/Bucher/Predal:1980} In runs where 
crystallization took place, a large spike in the
specific heat appeared, corresponding to a step or latent heat in the enthalpy
curve. Before looking at the composition dependence of $T_g$, we notice that 
the temperature dependence of $c_P$ is quite similar in form to
 experimental specific heat curves of 
Zr$_{41.2}$Ti$_{13.8}$Cu$_{12.5}$Ni$_{10.0}$Be$_{22.5}$ reported by Busch et
 al.~\cite{Busch/Liu/Johnson:1995} and of fluorozirconate and tellurite glasses
reported by Lin and Navrotsky;\cite{Lin/Navrotsky:1997, Lin/Navrotsky:1998}
there is an increase in specific heat in the supercooled liquid region 
compared to the high-temperature liquid region. For the tellurites, Lin and
 Navrotsky identified the source of this as specific structural rearrangements
that take place in the liquid prior to the glass transition. We will see in the
next section what evidence there is for structural rearrangements in the Cu-Mg
supercooled liquid.

\begin{figure}
\epsfig{file=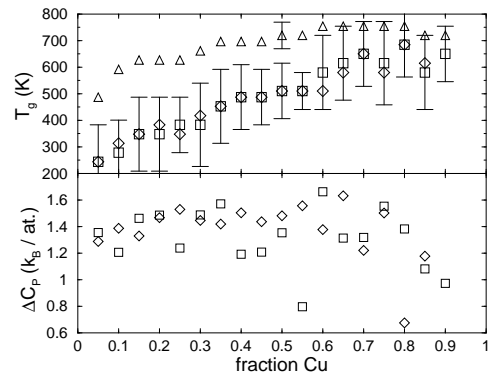,width = 2.5 in}
\caption{\label{TgSpHeatComp}$T_g$ and $T_{\textrm{split}}$ (upper panel) and
 $\Delta c_P$ (lower panel). Squares: $T_g$, $\Delta c_P$ at $0.72 K /ps$;
diamonds at $0.25 K/ps$. Triangles: $T_{\textrm{split}}$(Cu) at $0.72 K /ps$.}
\end{figure}

Fig.~\ref{TgSpHeatComp} shows $T_g$ and $\Delta c_P$, the heat capacity jump
(obtained by roughly determining the transition region as the peak in the 
derivative of $C_P$ and taking the difference of $C_P$ on either side of the 
peak) for different compositions 
and cooling rates. $T_g$ rises roughly linearly with increasing fraction of Cu,
which presumably reflects a general increase in energy scale as we go from
the weakly cohesive (low melting point) Mg to the more strongly cohesive Cu.
The fluctuations towards the Cu-rich end are due to the mid-point method's 
difficulty handling the somewhat less clean $c_P$-data there. The fluctuations
in $\Delta c_P$ are also due to the imperfect $c_P$-data. Nevertheless, it 
seems clear that $\Delta c_P$ has the value of roughly $1.5 k_B$ per atom, 
independent of concentration. This is a 
relatively small amount, which is typical of so-called ``strong'' 
glass-formers, which include most BMGs.\cite{Busch:2000} In particular the 
Mg$_{65}$Cu$_{25}$Y$_{10}$ shows a jump of the same order 
(actually $2k_B/\textrm{at.}$).\cite{Busch/Liu/Johnson:1998} However, we should
be careful about inferring strong-liquid behavior from this measurement; binary
alloys typically are not strong glass-formers,~\cite{Johnson:1999} and below 
we shall see evidence
of fragile-liquid behavior in the diffusivity. The apparent small jump of $C_P$
may be a consequence of the width of the transition.

\begin{figure}
\epsfig{file=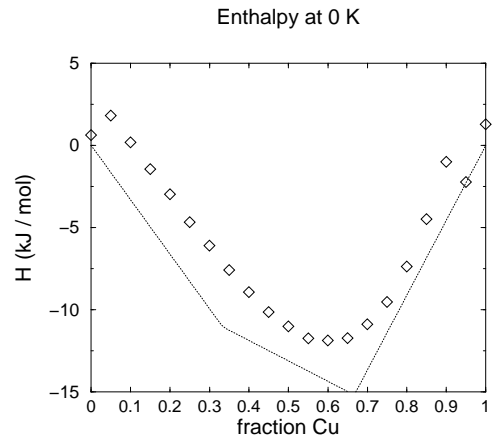,width = 2.5 in}
\caption{\label{formationEnthalpy}Diamonds: formation enthalpy per atom in
 final zero-temperature glassy state. Dotted line: formation enthalpy per atom
 of corresponding (in general two-phase) crystal.}
\end{figure}

As a partial means of determining how good, meaning how stable or
 well-annealed, the final configurations are, we consider their enthalpies. We
have seen already how the final enthalpy depends on cooling rate 
(Fig.~\ref{CoolingRateFig}); we now compare to the equilibrium phases, for 
different compositions. Fig.~\ref{formationEnthalpy} shows the {\it formation 
enthalpies} as a function
of composition. The formation enthalpy is the enthalpy minus the appropriate
linear combination of the pure elements' enthalpies. The appropriate quantity
 to compare to, also shown in Fig.~\ref{formationEnthalpy}, 
is the formation enthalpy of the corresponding crystalline phase, which in 
general is a two-phase mixture (so, e.g., between 33\% Cu and 66\% Cu it is
an appropriate weighting of the formation enthalpies of Cu$_2$Mg and Mg$_2$Cu).
We notice that the glass formation enthalpy follows quite closely the 
crystalline one, being 1--4 $kJ/mol$ higher (the exceptions being at 0, 
95\% and 100\% Cu where the system did in fact crystallize). This is quite 
small and typical of easy glass 
formers.\cite{Johnson:1986,Johnson:1999,Greer:1995} For the 15\%Cu composition
the value $4.2kJ/mol$ was reported by Sommer et al. for the transformation 
enthalpy from the
crystalline to the amorphous state, which is slightly higher than our value of
$3.53kJ/mol$---that is, our glass at this composition appears to be a little
too stable compared with experiment. This kind of discrepancy can only be due
to limitations of the interatomic potential, and not to the high cooling rate.
This gives us further confidence that we have created glassy structures which
 are more or less as stable as they can be.

\begin{figure}
\epsfig{file=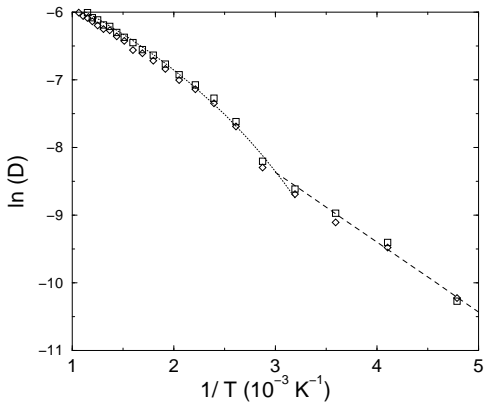,width=2.5 in}
\caption{\label{VFplotCu0.15} Diffusion constants in 15\% Cu. Squares: Mg; 
diamonds: Cu. Dotted line: VF fit to high temperature Mg data. Dashed line: 
Arrhenius fit to low temperature Mg data.}
\end{figure}

\begin{figure}
\epsfig{file=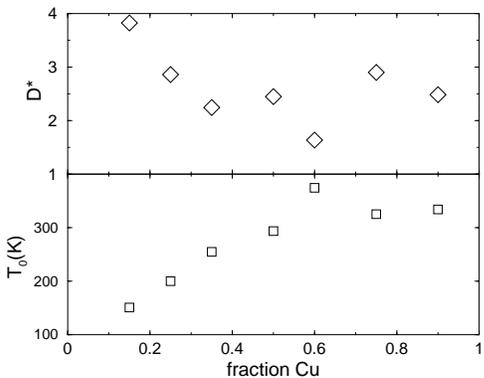,width=2.5 in}
\caption{\label{VFparameters}Upper panel: fragility parameter $D^*$ of 
Vogel-Fulcher fits to Cu diffusion constants at selected compositions. Lower
 panel: location of apparent divergence $T_0$ from the same fits.}
\end{figure}

For selected compositions and selected temperatures, configurations from the
cooling runs were used as initial configurations for further simulations in 
which diffusion constants for the two atomic species were measured. An 
Arhennius plot for the 15\% Cu composition is shown in Fig~\ref{VFplotCu0.15}. 
There is a clear indication of a transition near $1000/T \sim 3K^{-1}$, 
corresponding
 to $T \sim 330 K$, which is consistent with the $T_g = 350K$ obtained from the
specific heat. For each composition for which diffusion constants were 
measured, we have fitted the high temperature part of the data to the 
Vogel-Fulcher (VF) law

\begin{equation}
D = D_0 \exp (\frac{D^* T_0}{T - T_0})
\end{equation}

\nod where $T_0$ is the location of the apparent singularity and $D^*$ is the
so-called fragility parameter. In Fig~\ref{VFparameters} we show $D^*$ and 
$T_0$ obtained from fits of the Cu diffusion constants to the VF law (the Mg 
values are very similar, the differences being very small compared to the 
differences from composition to composition). There is
a reasonably clear trend towards decreasing $D^*$ and increasing $T_0$ as the 
fraction of Cu increases. High values $D^*$ are associated with strong 
glass-formers, the archetypal case of SiO$_2$ having $D^*=100$. Bulk metallic 
glasses are considered strong\cite{Busch:2000} with $D^* \sim 20$. So-called 
fragile glasses have $D^*$ around 2. From our diffusion
data we get low fragility parameters, in the range 2--4, indicating that the
Mg-Cu glasses are somewhat fragile. This is consistent with the experimental
fact that this is not in fact a bulk metallic glass. The $T_0$ values increase
as the fragility decreases, so that the apparent singularity approaches the 
actual
glass transition temperature. These trends, reflecting greater fragility 
(decreasing $D^*$) with increasing Cu composition, are also consistent with the
fact that experimentally, amorphous Mg-Cu can only be made at all for Mg-rich
 compositions, since strong liquids tend to be robust against crystallization 
(in a strong glass former the melt viscosity is high, making the kinetics 
slow). Thus, our diffusion results put the binary alloy Mg-Cu at the 
fragile end. This seems to contradict the suggestion of strong-liquid behavior
from the specific heat data. The small $\Delta C_P$, may have a 
simple explanation, however, namely that it has been reduced due
to the broadening of the transition in the simulations compared to what one 
would expect experimentally. This broadening implies that a certain amount of 
restructuring, that at slower cooling rates would take place above $T_g$, in
the simulation takes place during and below $T_g$. The net enthalpy change (the
 area under the $C_P$ curve) is more or less the same, so the height of the 
curve above $T_g$ must be reduced to compensate. Thus we can assert that the 
simulations are consistent with Mg-Cu being a fragile glass-former, like most 
binary alloys.

\section{\label{structuralAnalysis}Structural Analysis}

\begin{figure}
\epsfig{file=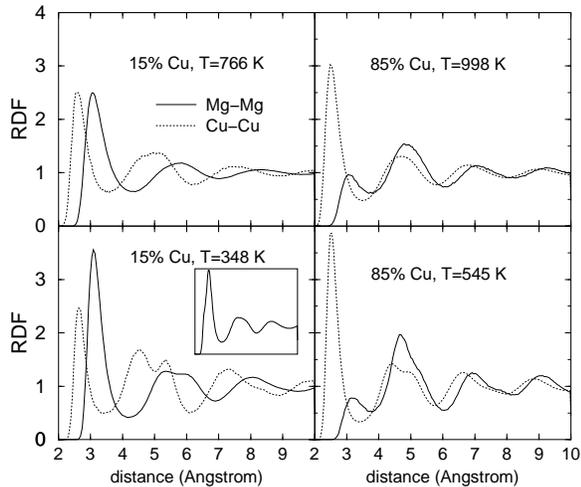,width=3.0 in}
\caption{\label{RDFplot23_4}RDFs, Mg-Mg and Cu-Cu. Left panels: 15\% Cu; right
 panels: 85\% Cu; upper panels: $T=T_{\textrm{eut}}$; lower panels: $T=T_g$;
 inset on bottom left panel: combined RDF (Mg-Mg, Cu-Cu, Mg-Cu).}
\end{figure}

\subsection{\label{radialDistributionFunction}Radial distribution function}

Fig.~\ref{RDFplot23_4} shows the partial RDFs $g_{\textrm{Mg-Mg}}(r)$ and 
$g_{\textrm{Cu-Cu}}(r)$ for two compositions at two temperatures. At
the higher temperature, which is the eutectic temperature for the corresponding
 region of the phase diagram, the system is expected to be in 
equilibrium, and the RDFs have the normal structure of a liquid, with nearest
neighbor distances of 3.1\AA\ for Mg and 2.6\AA\ for Cu, which are close to 
their values in the bulk crystal phases of the pure elements. The lower panels
in Fig.~\ref{RDFplot23_4} show the RDFs at the respective $T_g$ for each
composition. At 15\% Cu, the first peak is prominent for both RDFs. In the 
Cu-rich alloy on the other hand, the first Mg-Mg peak is significantly
 suppressed, indicating the Mg atoms are not particularly likely to be found 
next to each other. This is not surprising since we expect Mg-Mg bonds to be 
weak compared to both Cu-Cu and Mg-Cu bonds, given the cohesive energies of the
 pure elements and the intermetallic compounds.

We can see a distinct splitting of the second peak in
 $g_{\textrm{Cu-Cu}}$ in both compositions. The splitting occurs also for
 $g_{\textrm{Mg-Mg}}$, but at lower temperatures (here it is also obscured, 
particularly in the Cu-rich compositions, by the fact that the first sub-peak 
is significantly higher than the second, which thus appears as a shoulder on 
the high-side of the first). Such a splitting is commonly
associated with the glass transition, but we can see here that the splitting
is already well developed at $T_g$ for $g_{\textrm{Cu-Cu}}$ and in fact it
first occurs well above $T_g$. Fig.~\ref{TgSpHeatComp} shows the temperature
$T_{\textrm{split}}$ at which this occurs, determined in a somewhat arbitrary
 manner by visual inspection of the RDFs for different temperatures, as a
 function of composition. The
dependence on composition is rather less than that of $T_g$, and in fact it 
appears that the splitting is not related to the glass transition in a direct
 way. Note that what is typically observed experimentally---the combined RDF,
 which averages over the different components---does not show the splitting,
 because the location of the second peak differs for different 
components and the effect is washed out (see inset Fig.~\ref{RDFplot23_4}; in 
fact, for Mg-rich compositions, the total RDF has a split {\it
 first} peak due to the difference in location between Cu-Cu and Mg-Mg first 
peaks).

\begin{figure}
\epsfig{file=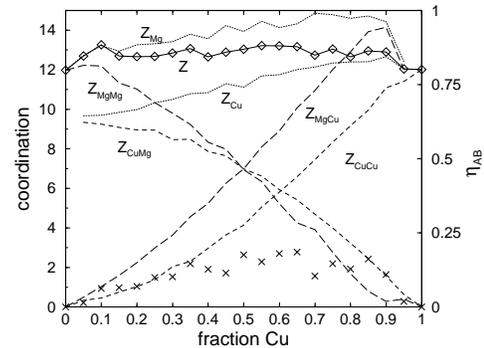, width = 2.5 in}
\caption{\label{coordinationNumbers}Partial and total coordination numbers as 
indicated. $Z_AB$ means the average number of neighboring $B$ atoms that an
 $A$ atom has. Crosses are Spaepen-Cargill short-range chemical order 
 parameter $\eta_{AB}$ determined from the coordination numbers. }
\end{figure}

\subsection{\label{coordinationNumbersSec}Coordination numbers}

By integrating the RDFs appropriately\footnote{That is, multiplying by 
$4\pi r^2 \rho_A$, where $\rho_A$ is the density corresponding to which type of
 neighbor is being counted, and integrating until the first minimum.} we can 
determine the partial, total and
 average coordination numbers, $Z_{AB}$, $Z_{A}$ and $Z$. These are shown in 
Fig.~\ref{coordinationNumbers}, for the zero temperature RDFs from the runs 
with the higher cooling rate ($0.72 K/ps$). We have checked that virtually
identical results are obtained with the lower cooling rate. The average 
coordination number is quite independent of composition, $Z = 12.91 \pm 0.17$.
The coordination number of Mg, $Z_{Mg}$, is always higher than the total
$Z$, and $Z_{Cu}$ always lower. Both rise as the fraction of Cu increases
 (their average does not because it is weighted by the concentrations). From 
the coordination numbers we can calculate the Spaepen-Cargill short-range order
 parameter~\cite{Cargill/Spaepen:1981}

\begin{align}
\eta_{AB} & = Z_{AB} / Z_{AB}^* - 1 \\
Z_{AB}^* & = c_B Z_A Z_B / Z
\end{align}

Here $c_B$ is the concentration of $B$ atoms (which we take as Cu). A positive 
value of $\eta_{AB}$ 
indicates a tendency for more unlike bonds than
would be expected in an  alloy which is completely chemically disordered. 
Fig.~\ref{coordinationNumbers} shows $\eta_{AB}$; it is definitely positive
throughout the glassy range of compositions, with an apparent maximum near the 
middle of the range. However while the maximum value $\eta_{AB}$ can ever take
is unity, the maximum for a particular composition is somewhat less, and values
should be normalized by the maximum before comparing different compositions.
We have not done this since it is not clear what $\eta_{AB}^{\textrm{MAX}}$
is in an amorphous system, particularly in the regime of $x_A \simeq x_B$ (see
Ref.~\onlinecite{Cargill/Spaepen:1981}).

\begin{figure}
\epsfig{file=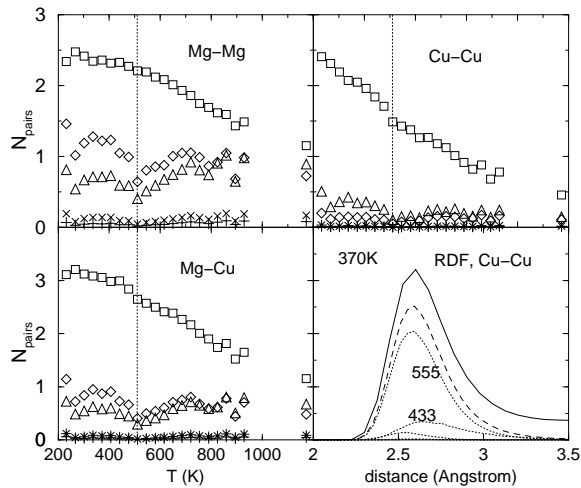, width= 3.0 in}
\caption{\label{FirstPeakAnalysis}Common Neighbor Analysis (CNA) of first peak
 of partial RDFs for
50\%Cu glass. Bottom right panel: RDF (solid line) contributions from 555, 544,
 433, 421 and 422 pairs (dotted lines), and the sum of these (dashed line). 
Other panels: number of neighbors of specified type (e.g. of a Cu atom which 
are Cu and make a 555 pair) as a function of temperature. Squares: 555, 
diamonds: 544, triangles: 433, plus: 421; cross: 422.}
\end{figure}

\subsection{\label{commonNeighborAnalysis}Common Neighbor Analysis}

To obtain more detailed information about the local atomic structure we use 
Common Neighbor Analysis 
(CNA).\cite{Jonsson/Andersen:1988,Clarke/Jonsson:1993} The analysis assigns
three indices to every pair of atoms, thus allowing a decomposition of the RDF
into contributions from different types distinguished by their CNA indices. The
first index is the number of neighbors that the two given atoms have in 
common; the second is the number of bonds among those neighbors, and the third
 is the size of the largest bonded cluster within the common neighbors (this
 last differs from the original definition,\cite{Clarke/Jonsson:1993} but 
agrees in all the cases of interest and is less ambiguous). The cutoff for two
 atoms to be considered ``neighbors'' or ``bonded'' is the position of the 
first minimum in the appropriate RDF. Note that the separation of the two atoms
 to whom the indices are assigned can be anything up to twice the nearest 
neighbor distance---beyond this, they cannot have any neighbors in common. 
Several groups have presented CNA analyses of the structure of
 metallic glasses.\cite{Qi/Wang:1991,
Clarke/Jonsson:1993, Posada-Amarillas/Garzon:1996, Moore/Al-Quraishi:2000, 
He/Ma:2001, Hui/others:2002} These all reported similar results: the first 
peak of the RDF is composed mostly of 555, 544 and 433 pairs, and the second
 peak is composed mostly of 333, 211 and 100 pairs. 555 pairs are associated 
with icosahedral order: in a perfect icosahedron the central atom makes a 555 
pair with each of its 12 neighbors. 544 and 433 pairs are formed when one or
 more bonds between the outer atoms of an icosahedron are broken. 333, 211 and
 100 pairs can also be associated with various pairs within a perfect 
icosahedron. Furthermore, the 333 and 211 pairs of the second peak combine to
 form the first sub-peak and the 100 
pairs make the second sub-peak, when the second peak splits.

In these papers the CNA was always performed on quenched configurations, 
obtained by rapid minimization to local minima from finite temperature 
configurations; this is preferable to doing the analysis on an instantaneous 
configuration at finite temperature, since the distortions caused by thermal 
fluctuations would in that case obscure the ``inherent structure''. Changes in
 the structure were correlated with the temperature from which the quench was
 made. In our analysis, we have taken an alternative approach to dealing with 
thermal fluctuations and have computed the full thermal averages of the 
contributions to the RDFs from pairs of different types. Analogous to the RDF
which is itself a thermal average, we thus obtain a ``radial distribution 
function'' for Cu-Cu/Mg-Mg/Mg-Cu pairs of type 555, 333, etc., which is in fact
an exact decomposition of the full RDF for the given species-pair. These
averages were computed during the same runs as the diffusion constants, with
starting configurations taken from the cooling runs at $0.72 K/ps$. The CNA
partial RDFs were computed every 10th major time step (starting with the 
20th---after which it was assumed the full RDF had converged sufficiently to 
read the position of the first minima). The CNA partial RDFs each consist of a
 single peak from which quantities such
 as peak position, height and width can easily be extracted. Also computed is
actual (average) number of pairs associated with such a peak, obtained by 
integrating the RDF against $4\pi r^2$ times an appropriate density. 
Furthermore we can see directly how these CNA-RDFs sum to give the full RDF 
for a given species-pair.

Fig.~\ref{FirstPeakAnalysis} shows the numbers of pairs in the sub-components
 of the first peak of the RDFs. We see the same broad picture described above,
 in terms of the roles played by 555, 544, 333, etc. pairs. This should not be
 surprising since as we shall see later icosahedral order is a dominating
 feature of the intermetallic alloys. In particular the 
number of 555 pairs grows more or less linearly as the temperature decreases 
from 1200 K to the glass transition temperature, beyond which it continues to 
increase, albeit with slightly smaller slope. The number of 421 and 422 pairs, 
associated with crystalline hcp and fcc order, is very small at all 
temperatures. The bottom right panel of Fig.~\ref{FirstPeakAnalysis} shows the
decomposition of the first peak of the RDF into contributions from the five
listed pair-types. The difference between the solid line (full RDF) and the 
dashed line (sum of the five listed pair-types) indicates that other types make
up a noticeable fraction. This were found to include small amounts of 311, 322,
 666, 533, and 532 pairs. At high temperatures when the number of 555 pairs is
low, all of these types of pairs, and some others not mentioned, contribute in
small amounts to make up the full coordination numbers. Thus the picture we
have of the liquid structure at high temperatures is one of many (we have seen
up to 15 different CNA types for nearest neighbor pairs) different 
local structures  constantly being created
and destroyed, and all contributing a little bit to the thermal average. As the
temperature cools, a pair of atoms is more and more likely to be found as a 555
pair. This is independent of what species the two atoms are, and of the
 compositions.

\begin{figure}
\epsfig{file=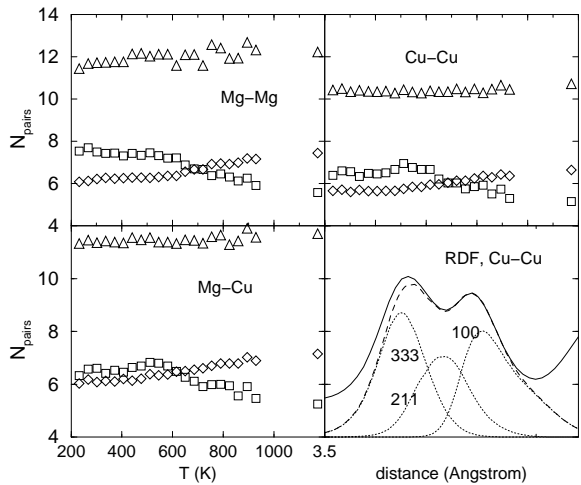, width= 3.0 in}
\caption{\label{SecondPeakAnalysis}CNA of second peak of partial
 RDFs for 50\%Cu glass. Bottom right panel: RDF (solid line) contributions from
 333, 211 and 100 pairs (dotted lines), and the sum of these three (dashed 
line). Other panels: number of neighbors of specified type (e.g. of a Cu atom
 which are Cu and make a 333 pair) as a function of temperature. Squares: 333,
 diamonds: 211, triangles: 100. In the Mg-Cu case, the number refers to Cu
 neighbors of Mg atoms.}
\end{figure}

Fig.~\ref{SecondPeakAnalysis} shows a similar analysis of the second peak. We 
see what others have found previously, that it is mostly made up of the
333, 211 and 100, and the first two making up the first sub-peak and the latter
 the second sub-peak. In fact there is only a small difference between the sum
of these three contributions and the full CNA, which appears on the 
shorter-distance side of the peak. This small difference is due to
 455, 444 and 322 pairs, which mainly occupy the region between first and 
second neighbor distances. At the highest temperatures (not shown), these last
three pairs make up a somewhat larger contribution, and are more clearly part
of the second (main) peak, but the 333, 211 and 100 are still dominant. The
numbers of pairs associated with these CNA-types change relatively little with
temperature: $N_{333}$ increases by about 30\% during cooling; $N_{211}$ 
decreases by the approximately same amount, leaving their sum constant 
($N_{100}$ is involved in this only to a small extent). It seems that 211 pairs
 are being transmuted to 333 pairs as the system cools. In the last section we
saw that the specific heat of the supercooled liquid is higher than the high 
temperature liquid, and noted how such behavior in experiments, termed 
``restructuring thermodynamics'',  has been associated with structural 
rearrangements that take place during cooling. In our system it is natural to
assume that the rearrangements identified by CNA analysis in this and the
preceding paragraph are responsible for the increased specific heat.


\begin{table*}

\begin{tabular}{|c|c|c|c|c|c|c|c|c|c|c|c|}
\hline
alloy & $A$ & $Z_A$ & $Z_{AMg}$ & $Z_{ACu}$ & $Z_{A444}$ & $Z_{A555}$ & 
$Z_{A666}$ & $N_{A333}$ & $N_{A211}$ & $N_{A100}$ & $N_{A455}$ \\
\hline
Cu$_2$Mg & Mg & 16  &   4 & 12  &  4  &  12 &  0  &  28  &  0   & 24   & 0 \\
         & Cu & 12  &   6 &  6  &  0  &  12 &  0  &  20  &  6   & 24   & 0 \\
{\it a}-Mg$_{0.35}$Cu$_{0.65}$ 
         & Mg & 15.7& 5.2 & 10.5& 2.1 & 9.8 & 2.6 & 13.9 & 12.4 & 22.9 & 3.6 \\
         & Cu & 12.6& 5.7 & 7.0 & 2.0 & 8.5 & 1.2 & 12.9 & 11.6 & 21.5 & 2.0 \\
Mg$_2$Cu & Mg & 15  &  11 &  4  &  0  &  12 &  3  &  22  &  7   & 24   & 2 \\
         & Cu & 10  &   8 &  2  &  2  &   8 &  0  &  16  &  4   & 26   & 0 \\
{\it a}-Mg$_{0.65}$Cu$_{0.35}$ 
         & Mg & 14.3& 9.6 & 4.7 & 2.3 & 10.1& 1.3 & 12.6 & 11.8 & 23.1 & 1.9 \\
         & Cu & 11.3& 8.7 & 2.6 & 2.2 & 7.9 & 0.5 & 11.0 & 11.1 & 21.2 & 1.0 \\
\hline
\end{tabular}
\caption{\label{intermetallicCNA}CNA figures for nearest ($Z$)
and next-nearest ($N$) neighbor pairs, for the intermetallic 
alloys Cu$_2$Mg and Mg$_2$Cu and amorphous alloys Mg$_{0.35}$Cu$_{0.65}$ and
Mg$_{0.50}$Cu$_{0.50}$. $Z_A$ and
 $Z_{AB}$ are partial and total coordination numbers. For the amorphous alloys
the figure under $Z_{A555}$ represents a sum over 555, 544 and 433 pairs.}

\end{table*}


\subsection{\label{comparisonOrderedStructures}Comparison with ordered
 structures}

At this point it is interesting to see what a common neighbor analysis of the 
intermetallic alloys Cu$_2$Mg and Mg$_2$Cu yields. The results are displayed 
in table~\ref{intermetallicCNA}, along with the partial and total coordination
 numbers. The numbers of different CNA types could be separated further 
by the species of the second atom, but the
table has already enough numbers. We see a distinct prevalence of 
nearest-neighbor pairs of type 555---almost all nearest neighbors are of this 
type, the rest being
444 and 666. It is impossible to have every nearest-neighbor pair being of the
555 type in a crystal, but it certainly seems that the crystal structures here
are trying to maximize the number of 555 neighbors. Now, ``icosahedral order''
strictly refers to having coordination number 12, all 555; however since in 
a binary alloy with a distinct size difference this coordination number is 
only achieved for the smaller atom, and only in a certain composition range,
strict icosahedral order cannot be attained, but we still choose to refer to
a high number of 555 pairs as representing ``icosahedral order''.

Of the second neighbor pairs only a few are of type 211, most being 333 and 
100. This is also consistent with icosahedral order: 333 pairs can be 
associated with pairs of tetrahedra which share a face, such as adjacent 
tetrahedra in a perfect icosahedron (or in the 555 structure, in view of our
 generalized sense of ``icosahedral''). 211 pairs
differ from 333 pairs by the removal of one of the common neighbors. It can
be supposed that the 211 pairs are defects of the icosahedral structure,
just as 544 and 433 pairs are, and thus that one would expect fewer of them 
relative to the number of 333 pairs in a more perfectly icosahedral structure.
This is consistent with the fact that the numbers of 211 pairs decreases as
temperature decreases in the glassy systems. In the same table are shown 
corresponding figures for the amorphous alloys of closest composition to the 
intermetallics, except that the numbers of 555, 544 and 433 pairs have been 
combined under the 555 column. The numbers for the amorphous alloys agree with
those from the corresponding crystalline phase to within 20 percent in most
cases, the biggest difference being the reduced number of 333 pairs, 
compensated more or less by the increase in 211 pairs. If we were to combine
the 333 and 211 figures, like we have the 555, 544 and 433 ones, we would see
that the structures in the amorphous and crystalline phases are locally very 
similar, the differences mostly being those between ``perfect'' 555 pairs and
``imperfect'' 544 and 433 pairs, and perfect and imperfect 333 and 211 pairs
respectively.

\begin{figure}
\epsfig{file=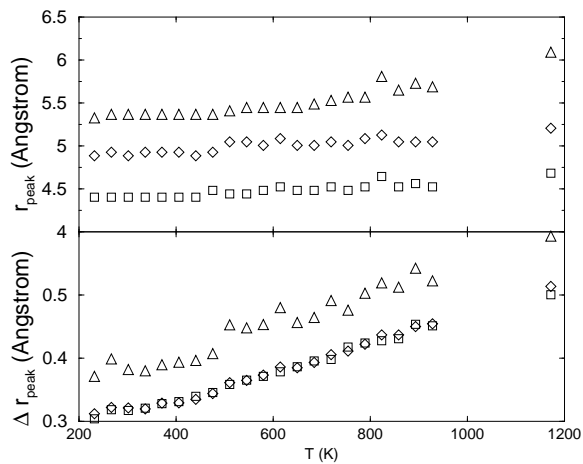, width = 3.0 in}
\caption{\label{PeakPosWidth}Positions (upper panel) and widths (lower panel) 
of CNA components of second RDF peak for 50\% Cu glass as a function of
 temperature. Squares: 333; diamonds: 211; triangles: 100.}
\end{figure}

\subsection{\label{explanationSecondPeak}Explanation of second peak splitting}

Our analysis indicates that the
contributions from various CNA-types vary smoothly with temperature. 
Fig.~\ref{PeakPosWidth} shows how the positions and widths of these peaks vary.
One expects the widths to decrease as temperature decreases, and this is indeed
the case. Their heights increase, mostly to compensate for the narrowing: we 
have already seen that the true measure of the weight of a peak, the number of 
pairs associated with it, has only a small temperature dependence in the case
of the second neighbor peaks. The splitting can now be seen as a natural 
consequence of this narrowing. It is also aided a little by the decrease in 
weight of the 211 peak, which is in the middle, and the corresponding increase
 of the 333 peak on the short side. Thus the 
splitting of the second peak does not itself indicate any kind of structural 
transition. It merely follows from the fact that the structure at this length
scale (second neighbor distance) associated with the liquid state remains as
 one cools into the glass state, and the narrowing of peaks which is to be 
expected as thermal motion decreases.

\section{Discussion and Summary}

Our main intent in simulating the cooling of Mg-Cu alloys has been to generate
glassy configurations that can be considered realistic enough for simulations
investigating mechanical properties. Mg-Cu is a first step towards the more
technologically interesting material Mg-Cu-Y. In order to assess the realism
of the simulations we have studied various aspects of the glass-forming nature
of the alloys: the thermodynamics, glass transition and structure. There are
three intrinsic limitations to these kinds of simulations: the interatomic 
potential, the system size and the timescale. We have reason to believe that
the EMT interatomic potential is not a major limitation in this study. We have
already discussed how much the physics of the binary amorphous alloy formation
 is based on size factors as well as the stability and structure of any nearby 
(in composition) intermetallic compounds. The fact that EMT parameters can be
chosen to match quite closely the formation enthalpies of the two Mg-Cu 
intermetallics means that the general bonding energetics are reasonably
well-represented. De Tendler et 
al.~\cite{DeTendler/Kovacs/Alonso:1992} applied the
 empirical model of Miedema for alloy formation to compute the glass-forming 
region of the Mg-Cu system. The close agreement with experiment they found 
indicates that there is nothing particularly unusual about this system.

This leads to the one feature of the Mg-Cu system which is poorly described by
our simulations: the extent of the glass forming region. The width of the 
glass-forming region is certainly a timescale-issue since the accessible
 timescales preclude nucleation of a 
crystal phase more complex than fcc; thus almost all compositions form a glassy
phase upon cooling. Issues of length scale could conceivably also be relevant 
for the formation of the more complex Mg$_2$Cu with its large unit cell. 
Of course, an advantage of being able to simulate glass-formation in a wide
 range of compositions is that it makes clearer that the splitting of the
 second peak and the glass transition are not coupled, since their dependences
 on temperature do not match. If one leaves aside crystallization, the
 fact that our results are 
largely independent of cooling rate, and the fact that the glass transition
temperature for the eutectic composition matches the experimental one, suggest
 that the timescale is not otherwise a problem---until the onset of the glass 
transition itself of course; there the fast cooling rates lead to the 
broadening of the transition compared to what would be expected 
experimentally. This of course does not rule 
out the possibility that there are relaxations that take place on  time scales
 significantly longer than those of our slowest cooling rate, yet still fast
 enough to take place during the experimental cooling. A possibility is that
such relaxations might be associated with a length scale larger than our system
size; thus our cooling rates are all slow enough to relax all structural 
rearrangements that are smaller than our system size, and thus we do not see 
any rate dependence, but perhaps we would see it in larger systems---the 
timescale issue and lengthscale issue would be in effect canceling each other
out. However any such extra relaxations must be very low energy, because the 
residual  enthalpies with respect to the crystal phases are as small as are 
measured experimentally. Another ``canceling'' possibility is that defects in
the interatomic potential, causing energy barriers to relaxation to be lower 
than they should, would lead to the relaxation times being lower and thus to 
the simulated cooling rate being more adequate than it otherwise should be.
Guerdane and Teichler~\cite{Guerdane/Teichler:2001} 
simulated Ni-Zr and ternary Ni-Zr-Al glass formation and obtained $T_g$s higher
 than experimental ones by a few hundred Kelvin, which they explained as being
due to the difference between their cooling rate ($10^{10}K/s$) and the 
experimental one, which makes it surprising that we do not see such a 
discrepancy. If it is indeed due to too-low relaxation barriers, this may not
matter so much for the purpose of obtaining low temperature glassy 
configurations; however it may be relevant for future studies of plastic 
deformation.

\begin{acknowledgments}
We thank Jim Sethna and Andrei Ruban for helpful discussion. This work was
 supported by the Danish Research Councils through
Grant No. 5020-00-0012 and by the Danish Center for Scientific
Computing through Grant No. HDW-1101-05. CAMP is sponsored by the Danish
 National Research Foundation.
\end{acknowledgments}

\bibliography{glass}

\end{document}